\documentclass[aps,groupedaddress,twocolumn,showpacs]{revtex4-1}
\usepackage[dvips]{graphicx}
\usepackage{dcolumn}
\usepackage{bm}
\usepackage{amssymb}
\usepackage{epstopdf}
\usepackage{amsmath}

\begin{document}

\title{Strong Electronic Correlation Originates from the Synergistic Effect of Large Moir\'e Cell and Strong Interlayer Coupling in Twisted Graphene Bilayer}

\author{Xun-Wang Yan$^{1}$}\email{yanxunwang@163.com}
\author{Jing Li$^{1}$}
\author{Yanyun Wang$^{1}$}
\author{Miao Gao$^{2}$}
\date{\today}

\affiliation{$^{1}$College of Physics and Engineering, Qufu Normal University, Qufu, Shandong 273165, China}
\affiliation{$^{2}$~Faculty of Science, Ningbo University, Zhejiang 315211, China.}

\begin{abstract}
By using the first-principles method based on density of functional theory, we study the electronic properties of twisted bilayer graphene with some specific twist angles and interlayer spacings. With the decrease of the twist angle(the unit cell becomes larger), the energy band becomes narrower and Coulomb repulsion increases, leading to the enhancement of electronic correlation; On the other hand, as the interlayer spacing decreases and the interlayer coupling becomes stronger, the correlation becomes stronger. By tuning the interlayer coupling, we can realize the strongly correlated state with the band width less than 0.01 eV in medium-sized moir$\rm \acute{e}$ cell of twisted bilayer graphene. These results demonstrate that the strength of electronic correlation in twisted bilayer graphene is closely related to two factors: the size of unit cell and the distance between layers. Consequently, a conclusion can be drawn that the strong electronic correlation in twisted bilayer graphene originates from the synergistic effect of the large size of Moir$\rm \acute{e}$  cell and strong interlayer coupling on its electronic structure.
\end{abstract}


\maketitle

As one of the most famous problems in condensed matter physics, the electronic correlation has always been a hot research subject.
Many transition metal oxides and rare earth metal compounds belong to the strongly correlated materials, in which the behaviors of electrons are complex due to the strong many-body interaction, bringing about lots of exotic physical properties, such as high temperature superconductivity, heavy fermion, and Mott insulator.
Generally, strongly correlated materials have partially filled d- or f-electron shells and there exists the strong Coulomb repulsion between d- or f-electrons. Meanwhile, the narrow and flat bands around the Fermi energy is the typical feature of their electronic band structures.

Apart from the materials containing metal element with d- or f-electron, some organic superconductor are known as strongly correlated materials.
The electronic correlation in Cs$_3$C$_{60}$, K-doped 1,2:8,9-dibenzopentacene (C$_{30}$H$_{18}$), (TMTSF)$_2$PF$_6$ and other organic superconductors are widely studied both in theory and experiment. \cite{Palstra1995,Xue2012,Lee2005}
Recently, twisted-angle bilayer graphene (TBG) with small 'magnic' angle of 1.1$^\circ$ exhibits the correlated insulating state and the intrinsic unconventional superconductivity, and the temperate-density phase diagram shows many similarities with that of high temperature cuprate oxide superconductors.
At small twist angle, the bands of TBG near the Fermi energy are flat and narrow.
These dipersionless bands are intimately associated to the electron correlation, and is regarded as the electronic signature of strongly correlated materials.
Seen from this perspective, the graphene bilayer superlattice, caused by the rotation of top layer with respect to bottom layer, provides a new scheme to create the strongly correlated state in the low-dimensional materials consisting of simple carbon element without d- or f-electron shells,
which can reduce the difficulty and complexity of the electronic correlation researches.

How to understand the electronic correlation in TBG systems is a vital issue, especially what factors result in the occurrence of strong correlation. The previous theoretical studies most forcus on the interaction of two Dirac cones states belonging to two graphene layers, which results in the merge of two Van Hove singularities in the case of magic angle.\cite{Trambly2012} There is few theoretical research to discuss the relationship of the electronic correlation and the interlayer coupling or Moir\'e cell size.
In this paper, by the first principles method, we investigate that the modulation of the spatial size and the strength of Moir\'e potential on the electronic properites of TBG system.

In our calculations, the plane wave pseudopotential method was used and implemented in Vienna Ab initio simulation package (VASP) program.\cite{PhysRevB.47.558, PhysRevB.54.11169} The generalized gradient approximation (GGA) with Perdew-Burke-Ernzerhof (PBE) formula~\cite{PhysRevLett.77.3865} as well as the projector augmented-wave method (PAW) \cite{PhysRevB.50.17953} for ionic potential were employed.
 The plane wave basis cutoff was set to 500 eV and the convergence thresholds for the total energy and force are 10$^{-5}$ eV and 0.005 eV/\AA ~.  The lattice parameter $c$ for the unit cell is set to 30 \AA~ to model the isolated bilayer graphene in $z$ axis direction. For the cell of TBG with a 6.01$^{\circ}$ angle, a mesh of $9\times 9\times 1$ k-points were sampled for the Brillouin zone integration.
Also, the van der Waals (vdW) interaction is included in our calculations.\cite{PhysRevLett.92.246401}

\begin{figure}
\begin{center}
\includegraphics[width=8.0cm]{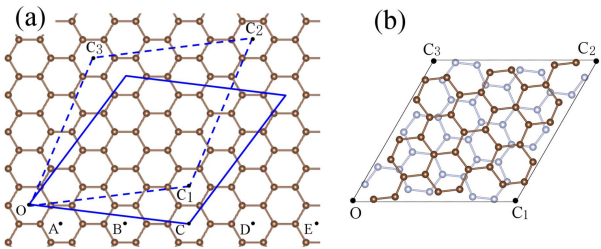}
\caption{Atomic structures of TBG with twisted angles of 21.79$^{\circ}$(a), 13.17$^{\circ}$(b), 9.43$^{\circ}$(c), and 7.34$^{\circ}$(d) are shown in 2$\times$2$\times$1 supercell. The numbers of atoms in one unit cell of TBG 28, 76, 148, 244 for the above four twisted angle cases. The top layer and bottom layer are blue and red, respectively.
  } \label{struct-cell}
\end{center}
\end{figure}

When one layer of graphene bilayer are rotated by an angle $\theta$, Moir$\acute{\rm e}$ pattern can be obtained.\cite{DeTramblyLaissardiere2010}
For a graphene layer, the Bravais lattice basis vectors are $\vec a_1$= ($\sqrt{3}$/2, -1/2)$a_0$, $\vec a_2$= ($\sqrt{3}$/2, 1/2)$a_0$.
A lattice vector $\vec V$ = m$\vec a_1$ + n$\vec a_2$ is rotated to $\vec {V^{\prime}}$ = n$\vec a_1$ + m$\vec a_2$ , the angle is defined as
\begin{equation}
cos(\theta) = \frac{n^2 + 4nm + m^2}{2(n^2 + nm + m^2)}
\end{equation}
the  Moir$\acute{\rm e}$ cell vectors are
$\vec t$= $\vec V$ = m$\vec a_1$ + n$\vec a_2$ and $\vec {t^{\prime}}$= -m$\vec a_1$ + (m+n)$\vec a_2$.
The rotation of one graphene layer is shown in Fig. \ref{struct-cell} (a). The lattice vector $\vec V$ = m$\vec a_1$ + n$\vec a_2$ is labeled as $\overrightarrow{OA}, \overrightarrow{OB}, \cdots$,
\begin{eqnarray}
\overrightarrow{OA}= 1 \cdot \vec a_1 + 0 \cdot \vec a_2; \qquad  \nonumber
\overrightarrow{OB}= 2 \cdot \vec a_1 + 1 \cdot \vec a_2; \\ \nonumber
\overrightarrow{OC}= 3 \cdot \vec a_1 + 2 \cdot \vec a_2;  \qquad
\overrightarrow{OD}= 4 \cdot \vec a_1 + 3 \cdot \vec a_2; \\ \nonumber
\overrightarrow{OE}= 5 \cdot \vec a_1 + 4 \cdot \vec a_2; \qquad
\overrightarrow{OF}= 6 \cdot \vec a_1 + 5 \cdot \vec a_2.  \nonumber
\end{eqnarray}
The vector $\overrightarrow{OA}$ (or $\overrightarrow{OB}, \cdots$) is one side of solid line rhombus, which denotes the size of one Moir$\acute{\rm e}$ cell. The solid line rhombus is rotated to the position of dash line rhombus and is superimposed on the bottom graphene layer to form the TBG.
Fig. \ref{struct-cell} (b) show the atomic structure of TBG unit cell with twist anglesof 13.17$^{\circ}$, in which the top layer is blue and bottom layer is red, and the atom number is 76 in such a unit cell.


\begin{figure}
\begin{center}
\includegraphics[width=8.0cm]{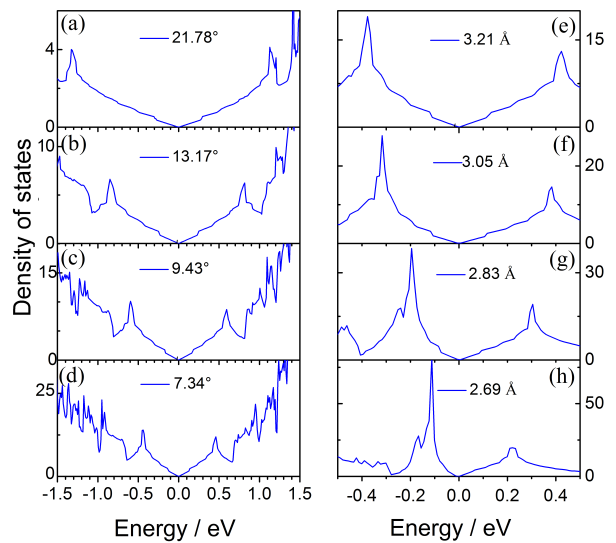}
\caption{Total DOS for TBG with the twist angles of 21.79$^{\circ}$, 13.17$^{\circ}$, 9.43$^{\circ}$, and 7.34$^{\circ}$, shown in (a), (b), (c), and (d) respectively. As the twist angledecreases, the peaks of DOS related to Van Hove singularities are more and more close to Fermi energy. For the TBG with $\theta$ = 7.34$^{\circ}$, the distance between two layers is decreased to 3.21 \AA, 3.05 \AA,  2.83 \AA,  and 2.69 \AA. The corresponding total DOS are shown in (e), (f), (g), and (h). With the distance being shorter, the DOS peaks are more close to Fermi energy.
   } \label{dos}
\end{center}
\end{figure}

The linear dispersion and Dirac cones in electronic bands are the important features of graphene.
For the TBG, the Dirac cones belonging to two layers interact each other to result in the sharp peaks of density of states (DOS) on both sides of Fermi energy.
\cite{Trambly2012}
We perform the electronic structure calculations for TBG with the twist angles of 21.79$^{\circ}$, 13.17$^{\circ}$, 9.43$^{\circ}$, and 7.34$^{\circ}$, and the total DOS are shown in Fig. \ref{dos}(a), (b), (c), and (d) respectively.
For these twist angles, the positions of sharp DOS peaks below Fermi energy are -1.33 eV, -0.84 eV, -0.59 eV, and -0.44 eV.
 As the twist angle decreases, the peaks of DOS
gradually move and approach the Fermi energy.
When the angle is less than 1$^{\circ}$,
the two Van Hove singularities at both sides of Fermi energy merge and form one sharp DOS peak at Fermi energy,
which is related to the instability of electronic states and lead to the occurrence of some novel physical phenomenons.

For a unit cell of TBG with the twist angle smaller than 2.0$^{\circ}$, the number of atoms is greater than 3000. From the viewpoint of first-principles calculation, it is difficult to simulate the electronic properties of such large TBG cell.
So, the first-principle study on TBG system 
with small twist angle
has rarely been reported.
However, we find out that
Van Hove singularity can be shifted to Fermi energy by compressing the spacing between two layers.
For the small TBG cell with $\theta$ = 7.34$^{\circ}$, the distance between two layers is compressed to 3.21 \AA, 3.05 \AA,  2.83 \AA,  and 2.69 \AA, and the corresponding DOS peaks below Fermi energy locate at -0.38 eV, -0.32 eV, -0.20 eV, and -0.11 eV, shown in Fig. \ref{dos} (e), (f), (g), and (h). 
Consequently, 
 the reduction of distance between two layers of TBG is a feasible approach to shift the Van Hove singularity to the Fermi energy.
Very recently, the role of the out of plane corrugation on the flat bands in TBG was investigated by Lucignano's group, and they pointed that the interplane distance changed about 0.2 \AA~ from AA to AB stacking region. \cite{Lucignano2019}
An important experiment demonstrated that at twist angle 1.27$^{\circ}$ larger than magic angle 1.1$^{\circ}$, suerconductivity was induced by imposing the hydrostatic pressure to vary the interplane distance.\cite{Yankowitz2019}
Our ideas are in line with these theoretical and experimental results.
Based on our calculations and recent researches, we conclude that in addition to twist angle, interlayer coupling is another key factor to tune the electronic correlation in TBG.

\begin{figure}
\begin{center}
\includegraphics[width=8.0cm]{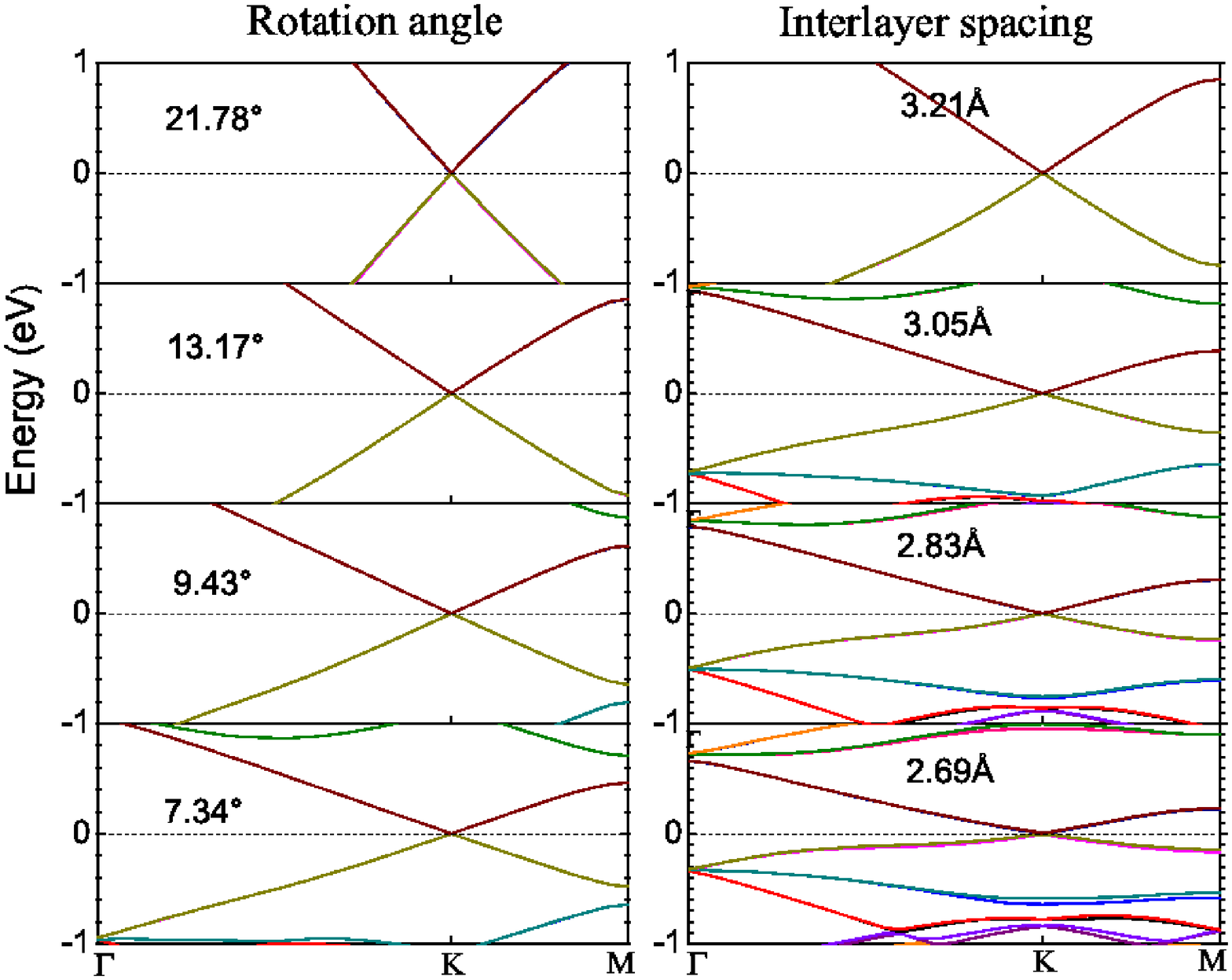}
\caption{Left panel: Band structures of TBG with the twist angles of 21.78$^{\circ}$, 13.17$^{\circ}$, 9.43$^{\circ}$, and 7.34$^{\circ}$. Right panel: Band structures of TBG with $\theta$ = 7.34$^{\circ}$ for the interlayer spacings of 3.21 \AA, 3.05 \AA,  2.83 \AA, and 2.69 \AA. The Fermi energy is marked by the dashed line. With the twist anglebeing smaller and the interlayer spacing being narrower, the valence and conductance bands become flatter.
   } \label{band}
\end{center}
\end{figure}

The band structures of TBG with the twist angles of 21.78$^{\circ}$, 13.17$^{\circ}$, 9.43$^{\circ}$, and 7.34$^{\circ}$ are shown in Fig. \ref{band} left panel.
With twisted angle decreasing, the size of graphene sheet in a unit cell become larger and the energy bands become narrower.
On the other hand, we fix the twist angle at 7.34$^{\circ}$, and compress the interlayer spacing to 3.21 \AA, 3.05 \AA,  2.83 \AA, and 2.69 \AA. The band structures of TBG with these reduced interlayer spacings are displayed in Fig. \ref{band} right panel.
It can be seen that the band width is reduced with the interlayer spacing being compressed.
In addition, with the twist angle and the interlayer spacing decreasing, the gap at M point between valence and conductance bands becomes smaller, while the Dirac cone at K point still persists, as shown in Fig. \ref{band}.
The width of those conductance bands for above situations are listed in Tab. \ref{Tab1} and Tab. \ref{Tab2}.

The electronic correlation is a vital topic at the current stage of TBG research.
For the TBG with small twist angle, the flat low-energy bands appears in band structure and the electron kinetic energy is quenched, leading to a strongly correlated phase.
Upon electrostatic doping away from the insulating state, the superconducting state with multiple similarities to cuprate's are observed. \cite{Cao2018,Cao2018a}
The TBG reminds us of Cs$_3$C$_{60}$ and K-doped aromatics,
in which the main element is carbon and benzene ring is the basic structural unit, similar to TBG.\cite{Kubozono2015,Yan2018,Yan2019}
What's important is that they are demonstrated to be the strong correlation systems in experiment and exhibit the unconventional superconductivity. \cite{Ruff2013,Takabayashi2009}
In K-doped aromaics superconductor, the superconducting transition temperature $T_c$ shows a linear relativity to the number of benzene rings. $T_c$ increases from 5 K for K$_x$phenanthrene with three benzene rings to 18 K for K$_x$picene with five benzene rings, and up to 33.1 K for K$_x$dibenzopentacene with seven benzene rings.\cite{Kubozono2015}
Among the three kinds of carbon-based superconductors of magic angle TBG, Cs$_3$C$_{60}$ and metal-doped aromatics, there must exist the similarity of electronic correlation because of their structural similarity.
The large cell size, involved in the large spatial period of Bravais lattice, is their important feature which results in the narrow energy band in electronic structure.\cite{Kittel2004}

A simple method to estimate the Coulomb repulsion in molecular crystal was proposed by G. Brocks $et ~al.$ \cite{Brocks2004}, and the effective Coulomb repulsion was expressed as $U_{eff} = U_{bare}-U_{screen}$.
The Coulomb repulsion $U_{bare}$ for two charges on single molecule can be derived from the difference of the energies of the neutral, doubly and singly charged molecules, and $U_{screen}$ is the screening energy involved in the polarization of the molecule and the neighbored molecules in crystal.
In the subsequent researches, the method was widely used to study the electronic correlation in alkali-metal-doped pentacene and picene. \cite{Craciun2009,Kim2011,Giovannetti2011}
Especially, Giovannetti $et ~al.$ and Kim $et ~al.$ calculated the strength of electron correlation of K-doped picene, denoted by the ratio of the Coulomb repulsion $U$ and the width of bands $w$, and found that it is close to the ratio in Cs-doped C$_{60}$.\cite{Giovannetti2011,Kim2011}
We adopt the similar method to calculate the Coulomb repulsion $U$ in TBG, 
whose two sheets of graphene in a unit cell are regarded as two molecules.
When one unit cell is charged by two or four electrons, corresponding to the energy band half-filling or filling for the electron-doped TBG system,\cite{Cao2018a,Cao2018}
each graphene sheet in unit cell has one or two additional electrons.
We label the energies of neutral, one electron charged, and two electron charged graphene sheet as $E(m)$, $E(m^-)$, and $E(m^{2-})$,
then the Coulomb repulsion between two electrons on each graphene sheet can be expressed as $U = E(m) + E(m^{2-}) - 2E(m^-)$.
TBG system is continuous and infinite in the $xy$ plane, the additional electrons is uniformly distributed over the TBG plane. In $z$ direction, the TBG is isolated and we use the large lattice parameter $c$ of 30 \AA~ to model the isolated bilayer graphene. So that no dipole correction was needed for TBG system, which is different from the charged picene molecule which is surrounded by other molecules to produce the screening effect.
Hence, the above formula $U$ is used to estimate the Coulomb repulsion in TBG systems.

\begin{table}
\begin{center}
  \caption{Band width, Coulomb repulsion, and electron correlation strength in TBG with different twist angles. }
  \label{Tab1}
  \begin{tabular}{llll}
    \hline
    $\theta$($^{\circ})$&$w$(eV)&$U$(eV)&$U/w$ \\
    \hline
    21.78 &2.33 &0.01 &0.004  \\
    13.17 &2.09 &0.36 &0.17  \\
    9.43  &1.41 &0.65 &0.46  \\
    7.34  &1.04 &1.14 &1.10  \\
    \hline
  \end{tabular}
\end{center}
\end{table}

\begin{figure}
\begin{center}
\includegraphics[width=8.0cm]{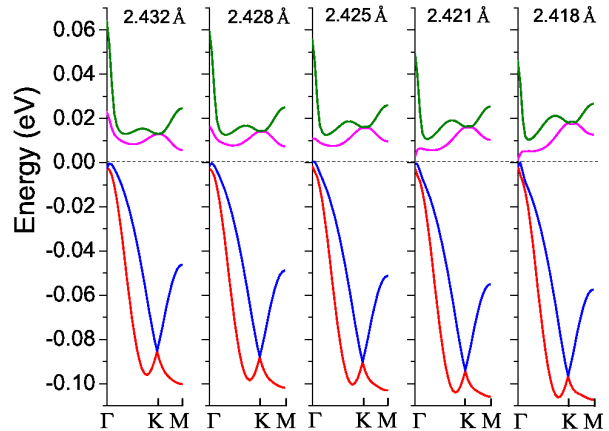}
\caption{The band structures of TBG with the
interlayer spacing of 2.432 \AA, 2.428 \AA, 2.425 \AA, 2.421 \AA, and 2.418 \AA. When the interlayer spacing changes very little, the electronic structure changes obviously. In middle panel, the width of band marked as magenta color is less than 0.01 eV.
} \label{band2}
\end{center}
\end{figure}

\begin{table}
\begin{center}
  \caption{Band width, Coulomb repulsion and electron correlation strength in TBG with different interlayer spacings. }
  \label{Tab2}
  \begin{tabular}{llll}
    \hline
    $d$(\AA)&$w$(eV)&$U$(eV)&$U/w$ \\
    \hline
    3.21 &0.99 &1.22 &1.23 \\
    3.05 &0.92 &1.25 &1.36 \\
    2.83 &0.79 &1.31 &1.66 \\
    2.69 &0.66 &1.35 &2.05 \\
    \hline
  \end{tabular}
\end{center}
\end{table}
We first inspect the influence of twist angle on the strength of electronic correlation in TBG systems.
Tab. \ref{Tab1} lists the data on the band width, Coulomb repulsion, and electronic correlation strength, which indicates that with twist angle decreasing the band width becomes smaller and Coulomb repulsion become larger, resulting in the enhancement of electronic correlation strength in TBG system. The result is consistent with the correlated electronic behavior of TBG at small angle observed in experiment.
Then, we examine the effect of the spacing between two graphene layers  on the strength of electronic correlation in TBG systems.
For the TBG with the angle of 6.01$^{\circ}$, the conduction band width, Coulomb repulsion, and the correlation strength are computed and presented in Tab. \ref{Tab2}. We can see that the electron correlation is enhanced greatly with the interlayer space decreasing.
As the twist angle of TBG decreases, the size of unit cell increases and the moir$\acute{\rm e}$ potential has the larger period in real space. Meanwhile, the reduction of interlayer spacing is related to the more strong interaction between two graphene layers.
The result can be roughly understood from Kronig-Penney model.\cite{Kittel2004} The TBG cell size and the interlayer coupling correspond to the distance between potential barriers and the height of potential barriers in Kronig-Penney model. With the barrier distance and barrier height increasing, the energy band will become narrower.
Based on the analysis and the data in Tab. \ref{Tab1} and \ref{Tab2}, we can draw a conclusion that the space period of moir$\acute{\rm e}$ potential and interlayer interaction, related to the cell size and interlayer spacing, are two main factors to determine the strength of electron correlation.
For TBG with magic angle, to simulate the flat band and van Hove singularity near Fermi energy is difficult because the unit cell contains too many atoms. 
However, if the interlayer spacing of TBG is reduced, we can use a medium-sized unit cell of TBG with narrow interlayer spacing to reproduce the flat band and van Hove singularity near Fermi energy.
Here, we choose the unit cell of TBG with twist angle of 6.01$^{\circ}$, which is composed of 364 atoms.
The band structures of TBG with the interlayer spacing of 2.432 \AA, 2.428 \AA, 2.425 \AA, 2.421 \AA, and 2.418 \AA~ are shown in Fig. \ref{band2}.
As can be seen, the width of conductance band marked as magenta color first decreases and then increases with the interlayer spacing reducing, and especially, the band width for 2.425 \AA~ situation in middle panel is less than 0.01 eV.
The results indicate that by carefully selecting the value of interlayer spacing we can obtain the flat and very narrow energy band of TBG.
 In such case, the Coulomb repulsion $U$ is about 1.24 eV, so its $U/w$ is larger than 100, which indicate that the strongly correlated state can be realized in medium-sized Moir$\rm \acute{e}$ cell of TBG by tuning its interlayer coupling.
In addition, it is worthwhile to emphasize that the medium-sized moir$\rm \acute{e}$ cell with strong correlation has a great significance for the superconductivity in TBG.
Generally, the superconducting transition temperature $T_c$ depends on the carrier density in strongly correlated superconductors.\cite{Cao2018}
For a charged medium-sized cell, the carrier density is much higher than the density of the Moir$\rm \acute{e}$ cell with magic angle.
So, we expect that the superconducting transition with high $T_c$ occurs in this medium-sized Moir$\rm \acute{e}$ cell of TBG.

In summary, we have performed the first principles calculations on the electronic properties of TBG with certain twist angles and interlayer spacings. The ratio of Coulomb repulsion $U$ and the energy band width $w$ is adopted to estimate the strength of electronic correlation of TBG systems. 
With the decrease of the twist angle and the interlayer spacing, the Coulomb repulsion becomes stronger and the band width become narrower, resulting in the enhancement of electronic correlation.
For a medium-sized cell of TBG, the reduction of interlayer spacing can result in the strong electronic correlation.
 These results indicate that the strength of electronic correlation in twisted bilayer graphene is closely related to two factors: the size of unit cell and the distance between layers. Consequently, a conclusion can be drawn that the strong electronic correlation in twisted bilayer graphene originates from the synergistic effect of the large size of Moire cell and strong interlayer coupling on its electronic structure.



We thank Professors Zhong-Yi Lu, Guo-Hua Zhong, Ping Zhang and Hai-Qing Lin for valuable discussions. This work was supported by the National Natural Science Foundation of China under Grants Nos. 11474004, 11404383, 11674087.

\bibliography{aa-2}

\end{document}